\documentstyle[12pt,epsfig,cite]{article}

\newlength{\dinwidth}
\newlength{\dinmargin}
\def\titlepage{\clearpage%
\setcounter{footnote}{0}\setcounter{page}{1}%
\thispagestyle{empty}\pagestyle{plain}\pagenumbering{arabic}%
\kern1mm\begin{center}
\end{center}
\vskip3mm\normalsize}
\def\docnum#1{\hbox to \hsize{\hskip123mm\hbox{#1}\hss}}
\def\date#1{\edef\@temp{#1}\ifx\@temp\@empty\def\@temp{\today}\fi
\hbox to \hsize{\hskip123mm\hbox{\@temp}\hss}}
\def\title#1{\vskip 0.8in plus 2in\begin{center}%
{\Large\bf#1\par}\vskip1.5em\end{center}\vskip 1in}
\def\@makefnmark{\hbox{$^{\@thefnmark)}$}}
\def\author#1{
\setcounter{footnote}{0}\def\@currentlabel{}%
\begingroup\def\thefootnote{\arabic{footnote}}
\def\@makefnmark{\hbox{$^{\@thefnmark)}$}}
\global\@topnum\z@ \large\begin{center}{\lineskip.5em
\begin{tabular}[t]{c}#1\end{tabular}\par}
\end{center}\par\vskip1.5em\@thanks\endgroup}

\def\abstract{\vskip0.8in plus 3in\begin{center}{\large\bf Abstract}\end{center}\quotation}

\newcommand{\gs}   {\gamma_{\rm S}}

\newcommand{\QGz}  {{\bf{Q}}^0}

\newcommand{\ls}   {\lambda_{\rm S}}
\newcommand{\ppb}  {$\rm{p\bar{p}}\;$}
\newcommand{\ee}   {${\rm e}^+{\rm e}^-$}

\newcommand{\jth}  {$j^{\rm th}$}

\newcommand{\dint} {{\rm{d}}}

\newcommand{\phiv} {\mbox{\boldmath $\phi\!$ \unboldmath}}
\newcommand{\ssb}   {$\rm{s\bar{s}}\;$}
\newcommand{\uub}   {$\rm{u\bar{u}}\;$}
\newcommand{\ddb}   {$\rm{d\bar{d}}\;$}

\newcommand{\qj}   {{\bf{q}}_j}

\newcommand{\E}    {{\rm{e}}}
\newcommand{\I}    {{\rm{i}}}

\begin{document}

\begin{titlepage}
\flushright{DFF 326/09/1998}
\flushright{September 1998}
\vspace{2.0cm}

\title{Strangeness counting in high energy collisions}

\centerline{\large{F. Becattini}}
\vspace{0.5cm} 
\centerline{\it{Universit\`a di Firenze and INFN Sezione di Firenze}}  
\centerline{\it{Largo E. Fermi 2, I-50125 Firenze (Italy)}} 
\centerline{e-mail: becattini@fi.infn.it}
\vspace*{1cm} 

\begin{abstract}
The estimates of overall strange quark production in high energy \ee, pp 
and \ppb collisions by using the statistical-thermal model of hadronisation are 
presented and compared with previous works. The parametrization of 
strangeness suppression within the model is discussed. Interesting 
regularities emerge in the strange/non-strange produced quark ratio which
turns out to be fairly constant in elementary collisions while it is
twice as large in SPS heavy ion collision.
\end{abstract}

\vspace*{1.5cm}
\centerline{\it{Talk given at Strangeness in Quark Matter 98 conference}}
\centerline{\it{July 20-24 1998, Padova (Italy), submitted to J. Phys. G}}

\end{titlepage}

\section{Introduction}

The interest for strangeness production in high energy collisions has
been strengthened by the possibility of its use as a probe of the Quark
Gluon Plasma formation \cite{raf}. An enhanced strange particle production
in heavy ion collisions with respect to nucleon-ion collisions have actually
been observed \cite{ody,wa}. In order to understand 
the relevance of such observations it is very important to study strangeness 
production in high energy elementary collisions like \ee or pp, as they provide
a natural baseline to gauge the results in more complex systems.\\
The comparison between strange and non-strange particle production in whatever 
kind of collision can hardly be accomplished by using only experimental data
without any input from models. In fact, not all particle species yields are 
practically measurable; moreover, even if an experiment was able to measure all 
stable (disregarding weak decays) hadrons species, strong decays following 
the primary 
hadron production modify the number of u, d, and (to a less extent) s valence 
quarks, thereby affecting the quantitative estimation of their actual production
in the hadronisation process. The direct experimental 
measurement of strange to non-strange quark production therefore would require
the determination of the inclusive yields of all hadron species, an exceedingly 
difficult experimental task.
As a consequence, a model of hadronisation is necessary to estimate the 
production of a large number of unmeasured hadrons and resonances. Of course
a prerequisite of any estimate is the agreement between the model and the
existing data with a fairly small number of free parameters in order not 
to spoil its predictive power.\\
In this paper I will discuss strangeness production by using a statistical-thermal 
model of hadronisation which turned out to be in good agreement with high 
energy collision data by using only three free parameters \cite{beca1,beca2}.       
  
\section{Statistical model and strangeness suppression}

The statistical-thermodynamical model of hadronisation has been described 
and discussed in detail elsewhere \cite{beca2,sudaf}. The main idea of the model
is the existence of a set of pre-hadronic clusters (or fireballs) converting
into hadrons according to the equiprobability of multi-hadronic phase 
space states (Gibbs postulate), where phase space is {\em locally} defined by 
the mass and the rest frame volume of each cluster. This assumption entails
the possibility of expressing overall hadron multiplicities (i.e. measured
in full phase space) by means of thermodynamical formulae within a canonical 
framework, that is with exact quantum number conservation.\\
The model, whose basic parameters are the temperature $T$ of the clusters/fireballs 
and the sum $V$ of their proper volumes, is supplemented with an extra 
strangeness suppression factor, which is formally beyond a pure statistical 
hadronic phase space model. This factor $\gs$ is 
introduced in the partition function used to derive hadron mulitplicities,
as a fugacity related to the valence strange quark content of the hadron 
\cite{gs}. The canonical partition function supplemented with $\gs$ reads 
\cite{beca2}:

\begin{equation} \label{1}
 Z(\QGz)= \frac{1}{(2\pi)^5} \int_{0}^{2\pi} \, \dint^5 \phi \,\, \E^{\,\I\, \QGz \cdot \phiv}
   \exp \,[V \sum_j \frac{(2J_j+1)}{(2\pi)^3}
       \! \int \dint^3 p \,\, \log \, (1 \pm \gs^{s_j} 
      \E^{-\sqrt{p^2+m_j^2}/T -\I \qj \cdot \phi})^{\pm 1}] \; .
\end{equation} 
where $\QGz = (Q, N, S, C, B)$ is a five-dimensional vector containing the initial
electric charge $Q$, baryon number $N$, strangeness $S$, charm $C$ and beauty $B$;
$J_j$ and $m_j$ are the spin and the mass of the \jth hadron species; $\qj$ is its 
quantum number vector $(Q_j, N_j, S_j, C_j, B_j)$ and $s_j$ its number of valence
strange+antistrange quarks. The hadron average multiplicities 
resulting from eq.~(\ref{1}), in the Boltzmannian limit, reads:

\begin{equation} \label{2}
  <n_j> \, = \gs^{s_j} (2J_j+1) \, \frac{VT}{2\pi^2} \, m_j^2 \, 
 {\rm{K}}_2(\frac{m_j}{T})\frac{Z(\QGz-\qj)}{Z(\QGz)} \; ,
\end{equation}
where the ratio $Z(\QGz-\qj)/Z(\QGz)$ is defined as {\em chemical factor}
\cite{beca2} and, unlike an ordinary fugacity, is not an intensive quantity
for it depends on the system size.\\
This way of regarding $\gs$ is in fact a grand-canonical one, as fugacities 
can be defined only in a grand-canonical framework. On the other hand, it is 
possible to define $\gs$ in a more general way which is independent of the 
adopted statistical formalism, i.e. microcanonical, canonical or grand-canonical.\\ 
Indeed, for a certain multihadronic state $\{n_1,\ldots,n_K\}$ ($n_1$ is the
number of hadrons belonging to species 1, ..., $n_K$ is the number of hadrons 
belonging to species $K$) originating from the collision, its phase-space 
probability is multiplied by a factor $\gs$ powered to the number of 
strange+antistrange quarks whose creation out of the vacuum is needed in 
order to set up that state. According to this definition, in a canonical 
framework, the probability $P$ of realizing a multihadronic state 
$\{n_1,\ldots,n_K\}$ in a collision whose initial state does not have any 
strange quarks is:

\begin{equation} \label{3}
 P \propto \exp(-E/T) \,\, \gs^{\sum_{j=1}^K n_j s_j} \, 
 \delta_{\sum_j n_j \qj,\QGz} \; , 
\end{equation}
where $s_j$ is the number of valence strange+antistrange quarks contained in 
the \jth hadron. This probability can be worked out to calculate the canonical 
partition function leading to the same expression as in eq.~(\ref{1}). The 
above definition can be easily extendend to a microcanonical or a grand-canonical 
framework. Furthermore, it should be emphasized that this definition is more 
general and more appropriate for collisions with initial strange quarks (for 
instance K p) in which the use of the same partition function (\ref{1}) obtained 
for colliding systems devoid of valence strange quarks would lead to odd results. 
In this special case, the probability (\ref{3}) becomes:

\begin{equation} \label{4}
 P \propto \exp(-E/T) \,\, \gs^{\sum_{j=1}^K n_j s_j - |S|} \, 
 \delta_{\sum_j n_j \qj,\QGz} \; , 
\end{equation}
where $|S|$ is the initial absolute strangeness, so that the canonical partition
function will differ by a factor $1/\gs^{|S|}$ from the (\ref{1}).

\section{Results on strangeness production}

The free parameters ($T$, $V$ and $\gs$) of the statistical-thermal model are 
determined by a fit to the measured hadron multiplicities (or multiplicity ratios)
by taking into account the decay chain following the primary production. 
The quality of the fits is good \cite{beca2} and the multiplicity of up to 25-30
particles can be well reproduced. For heavy ion collision\cite{becah} one more free 
parameter is necessary (the baryon-chemical potential) as the number of participant 
nucleons, i.e. the total baryon number, is a measured quantity unlike in \ee, pp 
and \ppb where it is known {\it a priori}. 
\begin{table*}
\caption[]{Results of thermal model fit to particle ratios at AGS Si+Au collisions.
The number of participant nucleons has been set to 108 and electric charge to 45
according to the estimation of ref.~\cite{satz}. The fitted parameters are 
$T$, $V T^3 \exp[-0.7 {\rm GeV}/T]$, $\gs$ and $\mu_B/T$ whilst $\ls$ is 
derived from the fitted parameters and their errors. The left column is the
set of parameters obtained by setting the error on the number of participants to
10, the right column to 0.1 and keeping the same total electric charge. Besides 
experimental errors, also the uncertainties on masses, widhts and branching ratios 
of hadrons have been taken into account according to the procedure described in 
ref.~\cite{becah}. The results are consistent with those in refs.~\cite{pbm,satz}.}
\begin{tabular}{lll}
\noalign{\smallskip} \noalign{\smallskip}  
\hline\noalign{\smallskip}   
 & Fit 1 & Fit 2  \\
\hline\noalign{\smallskip}
 $T$ (MeV)                         &  118.4$\pm$11.6   &  132.8$\pm$ 11.3  \\
 $V T^3 \exp[-0.7 {\rm GeV}/T]$    &  0.73$\pm$0.25    &  1.04$\pm$0.25    \\
 $\gs$                             &  0.923$\pm$0.16   &  0.792$\pm$0.12   \\
 $\mu_B/T$                         &  4.41$\pm$0.49    &  3.89$\pm$0.34    \\
 $\chi^2/dof$                      &  21.1/6           &  22.5/6           \\
\hline\noalign{\smallskip}
 $V$ (fm$^3$)                      &  $\approx 1250$   &  $\approx 664$    \\ 
 $\ls$                             &  0.488$\pm$0.093  &  0.569$\pm$0.10   \\ 
\hline\noalign{\smallskip}   
\end{tabular}
\end{table*}
The extracted $\gs$ values in elementary collisions at various centre of mass 
energy points are shown in fig. 1 along with those for heavy ion collisions at
SPS energies \cite{becah}; also shown the result of a fit to AGS Si+Au hadronic 
ratios quoted in ref.~\cite{satz} (see table 1). All $\gs$'s turn out to be 
definitely less than 1 except the AGS point which is consistent with 1, 
in agreement with the results of refs.~\cite{pbm,satz}, though with a large
uncertainty. This finding indicates that a strangeness chemical equilibrium at 
hadron level is not attained in most of the examined collisions. However, no 
apparent regularity emerges from the plot, as $\gs$ has different values in \ee 
and heavy ion collisions with respect to pp and \ppb collisions. The situation 
drastically changes when computing the ratio $\ls$:

\begin{equation} \label{5}
   \ls =\frac{\langle {\rm s}\bar{\rm s}\rangle} 
{0.5(\langle{\rm u}\bar{\rm u}\rangle + \langle{\rm d}\bar{\rm d}\rangle)} \; ,
\end{equation}
between newly produced valence \ssb pairs and half the sum of newly produced valence
\uub and \ddb pairs. As shown in fig. 2, $\ls$ has a pretty constant value of around 
0.2 for all examined \ee, pp and \ppb collisions whereas it is twice as high in 
heavy ion collision both at SPS and AGS energies; the central $\ls$ value at AGS is
indeed larger than at SPS but the fit error is so large that a conclusion is not
allowed. It should be emphasized that $\ls$ is a strangeness
suppression parameter related to the {\em quark content} of hadrons whereas $\gs$ 
is related to the strange hadrons phase space. The computation of $\ls$ proceeds
as follows: firstly, the best fitted primary average multiplicities $<n_j>$ for all 
hadron species are used to count the total number of quarks $<Q_i>$ of a given
flavour $i$:

\begin{equation} \label{6}
 <Q_i> = \sum_j <n_j>(T,V,\gs) \,\, q_{ij}
\end{equation}
where $q_{ij}$ is the number of valence quarks of flavour $i$ contained in the
\jth hadron. Secondly, the initial colliding quarks are subtracted so to count
only the primarily newly produced ones. The errors on $\ls$ are estimated by 
propagating the fit errors on $T$, $V$ and $\gs$ onto the primary multiplicities 
$<n_j>$. Needless to say, the reliability of the estimate where only few 
hadron species were measured relies upon the agreement between model and data 
where a large number of measurements are available.\\
The reason for the different behaviour of $\gs$ and $\ls$ in elementary collisions
is twofold. On one hand, a hadron gas with fixed $\gs$, $T$ and $V$ has an 
increasing value of $\ls$ for an increasing total baryon number (see fig. 3); hence,
if $\ls$ is constant, $\gs$ must be lower in pp collisions (baryon number 2) than 
in \ee and \ppb, provided that temperatures and volumes have similar values as
it actually is \cite{beca2}. The
physical reason of the increase of $\ls$ with $B$ in a hadron gas is the lower energy 
threshold for strange pair production in a baryon-rich environment where 
$\Lambda + {\rm K}$ production is favoured in comparison with kaon pair production 
in a baryon-free environment. On the other hand, if $\ls$ is constant, a decrease
of $\gs$ with increasing $V$ is expected, due to the so-called canonical strangeness
enhancement \cite{cley,soll} implying an increase of the strange hadron chemical
factor (see eq.~(\ref{2})) in increasingly large systems (see fig. 4). This effect 
accounts for the $\gs$ difference between \ee and \ppb collisions as the multiplicities 
and the volumes
of the latter are somewhat higher for the considered centre of mass energy points. 
Other smaller contributions arise from the anticorrelation between $\gs$ and $T$ 
(outcoming temperatures in \ee are slightly lower in \ee than in \ppb) and from the 
relatively more abundant production of heavy flavours in \ee collisions which slightly 
affect the fit of $\gs$.\\
If $\ls$ is more fundamental than $\gs$, the best way of parametrising strangeness 
suppression would be to fix the {\em mean} absolute value of strangeness (with possible 
superimposed fluctuations) instead of using $\gs$. This would amount to extend the 
conservation laws from 5 ($Q,N,S,C,B$) to 6 ($Q,N,S,C,B,|S|$) in the calculation of 
canonical partition functions in eq.~(\ref{1}), where $|S|$ is now meant to be the overall 
number of strange quarks including both initial and newly produced ones. An interesting 
question, under current investigation, is why $\gs$ has worked well anyhow in reproducing 
the strangeness suppression. This is not surprising for large systemes because the demand 
of exact $|S|$ leads to a fugacity, namely $\gs$ (see for instance ref.~\cite{kapo}), 
but it is a relevant question for elementary collisions where this is not possible;
perhaps the replacement of $\gs$ with $\ls$ might lead to a further improvement of
the fits.\\   
The $\ls$ value estimated by the statistical-thermal model of hadronisation in 
elementary collisions is in agreement with a previous estimate quoted in ref.~\cite{wro} 
only for centre of mass energies $\sqrt{s}<100$ GeV, whereas its rise in \ppb 
collisions at $\sqrt s > 100$ GeV claimed there is not observed. The reason 
of this discrepancy is the fact that $\ls$ was estimated in ref.~\cite{wro} by using 
only K/$\pi$ ratio as experimental input and two parametrizations of hadron 
multiplicities \cite{AK,SS} which, unlike ours \cite{beca2}, do not satisfactorily 
reproduce all available measured multiplicities in \ppb collisions. As an example,
for \ppb collisions at $\sqrt s=$ 546 GeV, the parametrization in ref.~\cite{AK} 
predicts a $\Lambda$/K$^0_s$ ratio of 0.49, by taking the $\ls=0.28$ value quoted 
in ref.~\cite{wro}, whereas the experimental value is 0.24$\pm$0.05 \cite{ua5}.\\
The enhancement of $\ls$ in heavy ion collision at SPS energies is indeed very 
interesting. An explanation of this effect by advocating the formation of a fully
equilibrated hadron gas is not viable, as $\gs=1$ seems to be ruled out from
recent analyses of the presently available data \cite{becah,kapo,gore}. 
On the other hand, it is not possible either to explain the rise of $\ls$ by 
invoking a canonical strangeness enhancement from pp to heavy ion collisions
at fixed $\gs$ just because a fixed $\gs$ would not be able to account for the
production of $\phi$ meson in pp collisions \cite{soll}.\\ 
A fully equilibrated hadron gas seems to be able to account for AGS data at 
$\sqrt s = 5.4$ GeV, in agreement with the conclusions of refs.~\cite{pbm,satz}, 
though the outcome error on $\gs$ is quite large. However, two {\it caveats} for
AGS fit should be mentioned; firstly, the ratios used in the fit have been 
measured in limited rapidity intervals whereas the model calculations involve
full phase space multiplicities. Secondly, the fit has been performed within 
a pure grand-canonical framework, while a canonical treatment should 
bring about a slight $\gs$ rise because a strangeness suppression due to the 
finite volume is involved. However, for the large extracted $V$ values 
(see table 1), this is expected to be a minor correction.    

\section{Conclusions}

A statistical-thermal model of hadronisation supplemented with a non-equilibrium
strangeness suppression parameter $\gs$ has been used to study primary strangeness 
production in elementary and heavy ion collisions at high centre of mass energy.
The reliability of the model is based on its success in reproducing many particle
multiplicities at several centre of mass energies, with only three free parameters. 
It is found that the ratio of newly produced strange to non strange quark 
$\ls$ (see eq.~(\ref{5})) is fairly constant for all elementary collisions.
This constancy emerges as an universal feature of hadronisation along with the
constancy of temperature \cite{beca2}. Thus, while u and d quarks are produced in 
the same amount, strange quarks are produced in a fixed ratio with u, d, which 
is most likely related to their higher mass. All quark pairs generated before or 
during the hadronisation process eventually fill hadronic phase space according
to a pure statistical law. Hence, the fact that $\gs < 1$ in a modified hadron
gas framework, simply reflects the underlying more fundamental $\ls \simeq 0.2$ 
parameter. In order to emphasize the role of the hadronisation model in extracting 
the $\ls$ regularity, it should be noticed that pure experimental observables such 
as the K$^+$/$\pi^+$ ratio does not show up such constancy (see fig. 5). In fact, 
kaon yield is enhanced in \ee collisions with respect to the examined pp and \ppb 
collisions because of the contribution of heavy flavoured hadron decays.\\
In heavy ion collisions at SPS energies, $\ls$ turns out to be twice as large.
This enhancement cannot be accounted for a hadron gas explanation and it is 
a subject of reflection. There have been some attempts to explain it in terms of
Quark-Gluon Plasma formation \cite{soll,gazd}.           

\section*{Acknowledgements}

I warmly thank the organizers of the {\it Strangeness in Quark Matter 98}
conference for their encouragement and for having provided a stimulating
environment. Stimulating discussion with J. Schukraft is gratefully 
acknowledged. 

\newpage



 \newpage
 
\section*{Figure captions}
  
\begin{itemize}

\medskip 

\item[\rm Figure 1]  
  Fitted $\gs$ in \ee, pp, \ppb and heavy ion collisions as a function
  of centre of mass energy (nucleon-nucleon centre of mass energy for heavy ions).
  
\item[\rm Figure 2]
 Estimated $\ls$ in \ee, pp, \ppb and heavy ion collisions as a function
 of centre of mass energy (nucleon-nucleon centre of mass energy for heavy ions).
 The two different estimates for \ppb collisions correspond to the two limiting
 cases of annihilation (lower) or survival (higher) of the 6 colliding valence 
 quarks.

\item[\rm Figure 3]
 The $\ls$ parameter in a hadron gas at fixed $T$, $V$ and $\gs$ 
 as a function of the number of initial protons (baryon number equal to electric charge).

\item[\rm Figure 4]
 Chemical factor $Z(0,0,1,0,0)/Z(0,0,0,0,0)$ of a neutral strange hadron
 (e.g. K$^0$) in a completely neutral hadron gas as a function of the volume for
 $\gs=0.5$ and three different values of temperature. The chemical factor approaches
 its grand-canonical limit, i.e. 1, for large volumes.
        
\item[\rm Figure 5]
 K$^+/\pi^+$ ratio in elementary collisions as a function of 
 centre of mass energy.
 
\end{itemize}

\newpage  
              
\begin{figure}[htbp]
\mbox{\epsfig{file=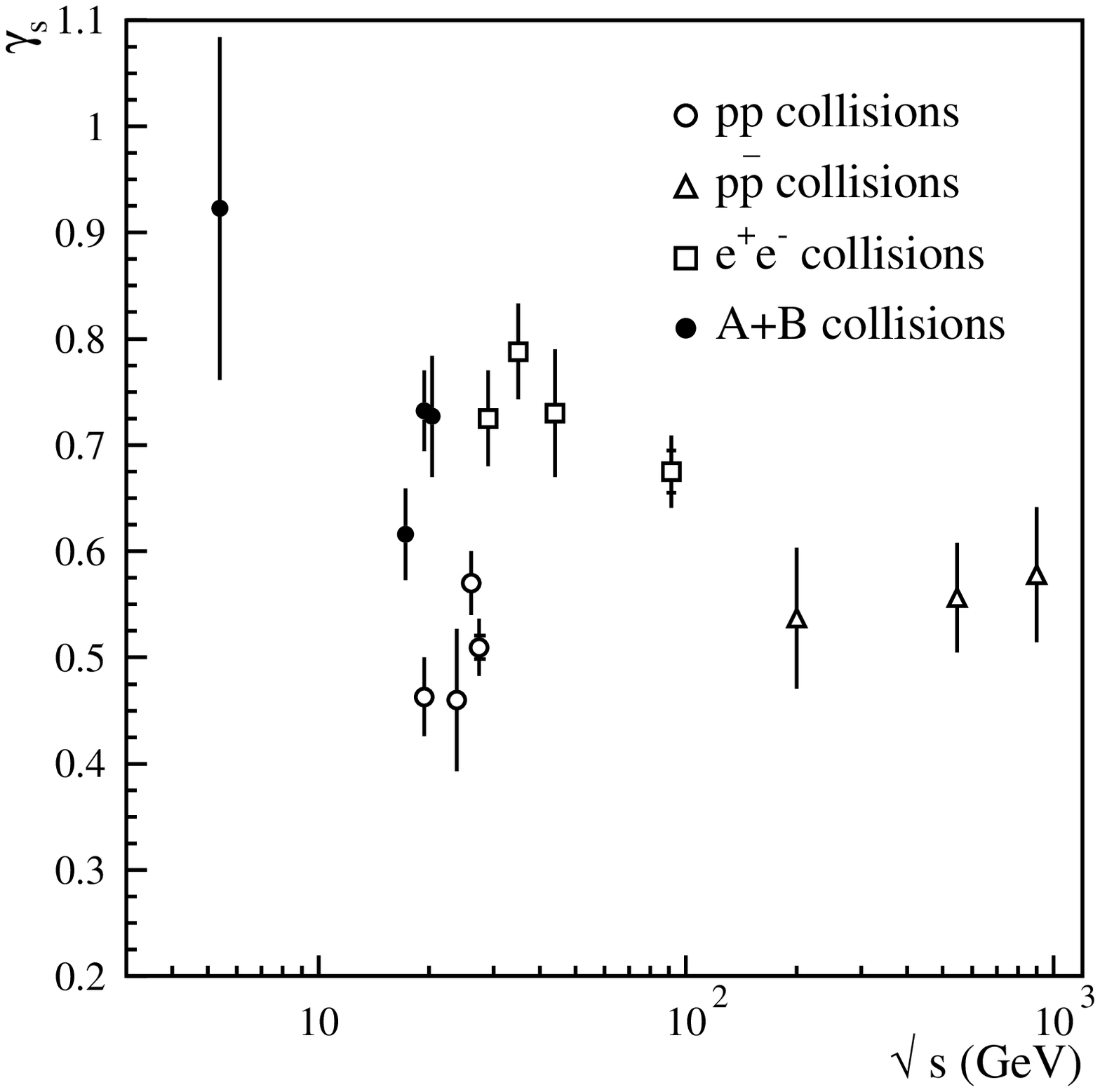,width=17cm}} 
\caption{}
\end{figure}

\newpage  
              
\begin{figure}[htbp]
\mbox{\epsfig{file=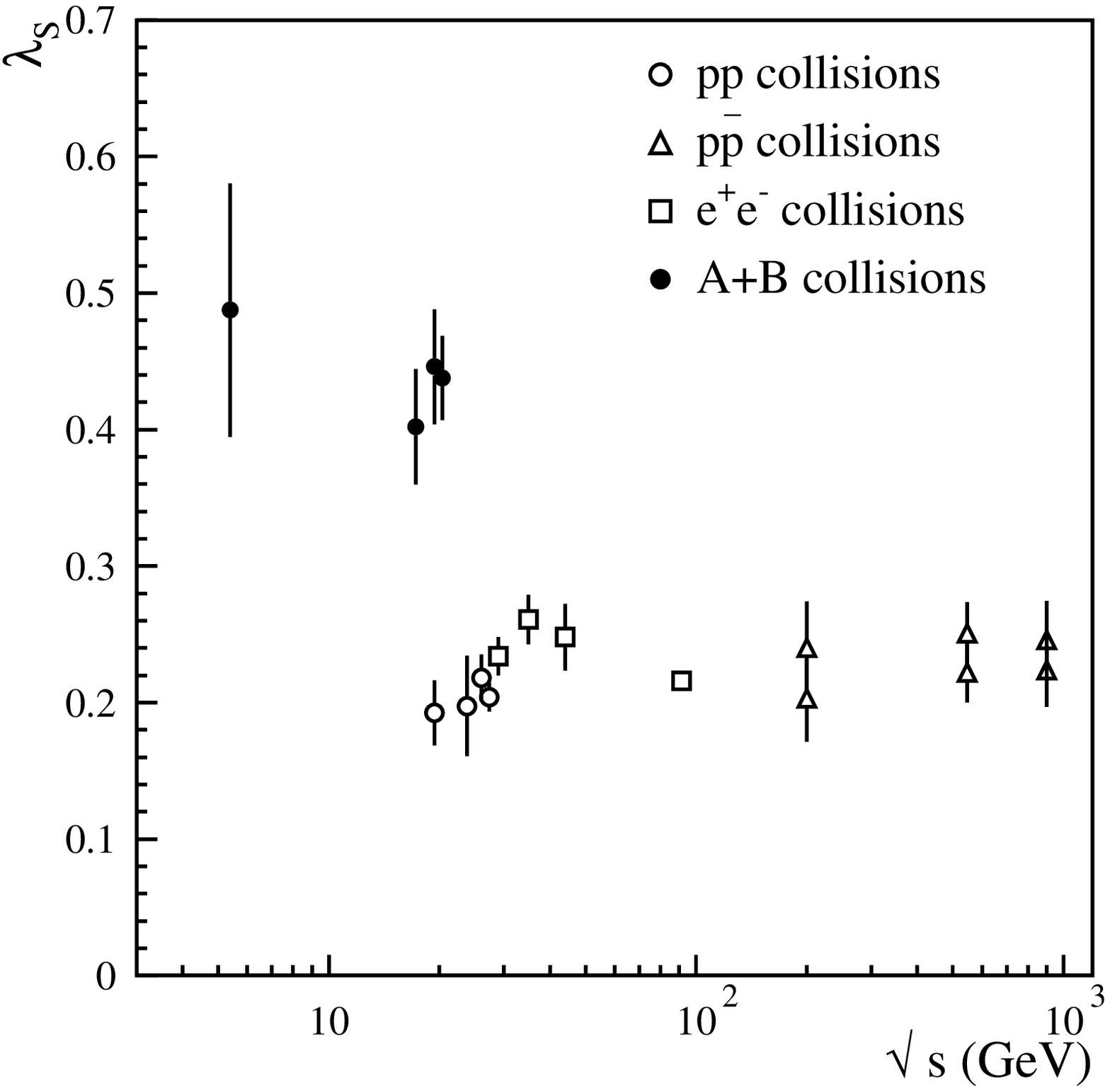,width=17cm}} 
\caption{}
\end{figure}

\newpage  
              
\begin{figure}[htbp]
\mbox{\epsfig{file=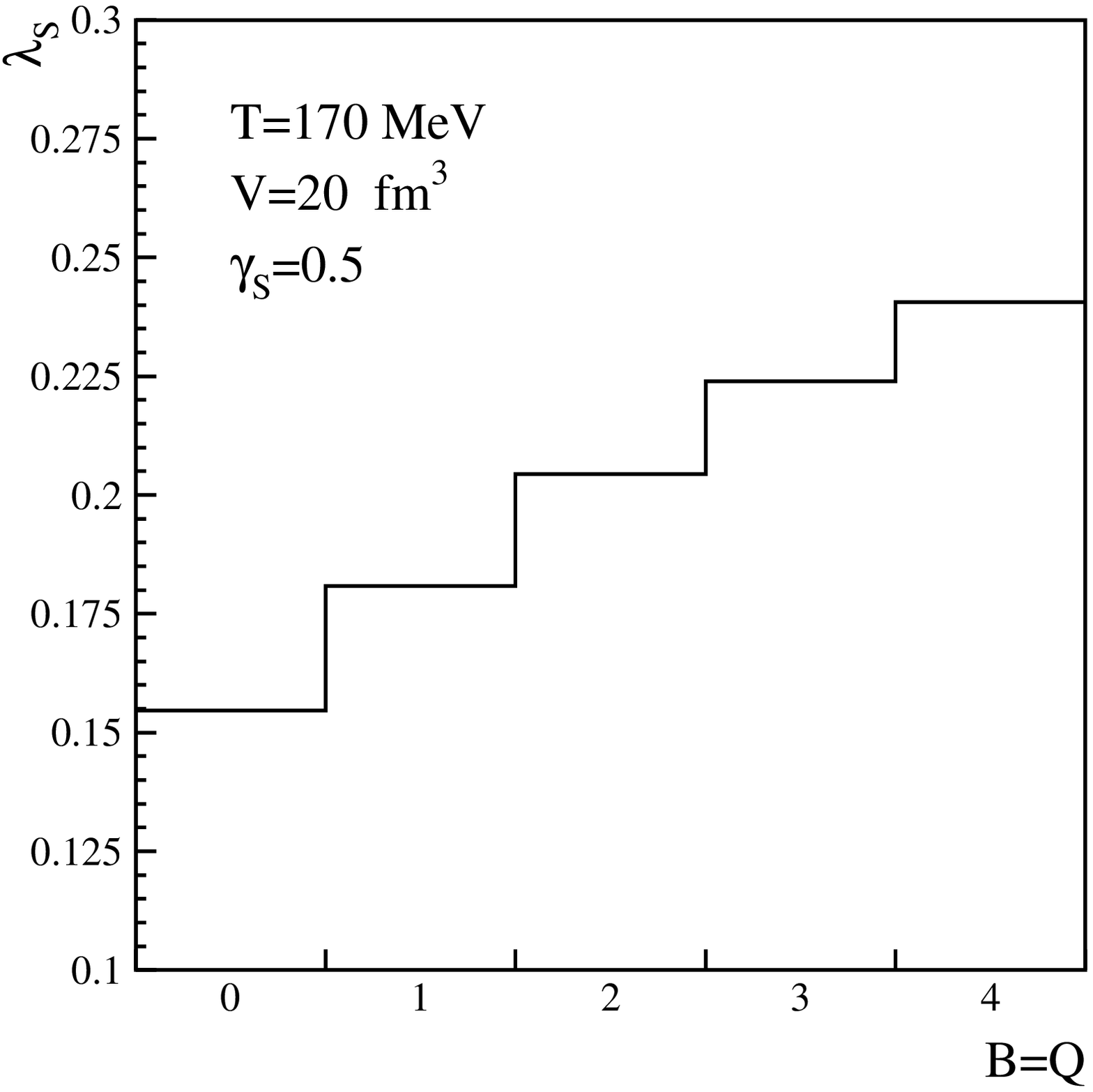,width=17cm}} 
\caption{}
\end{figure}

\newpage  
              
\begin{figure}[htbp]
\mbox{\epsfig{file=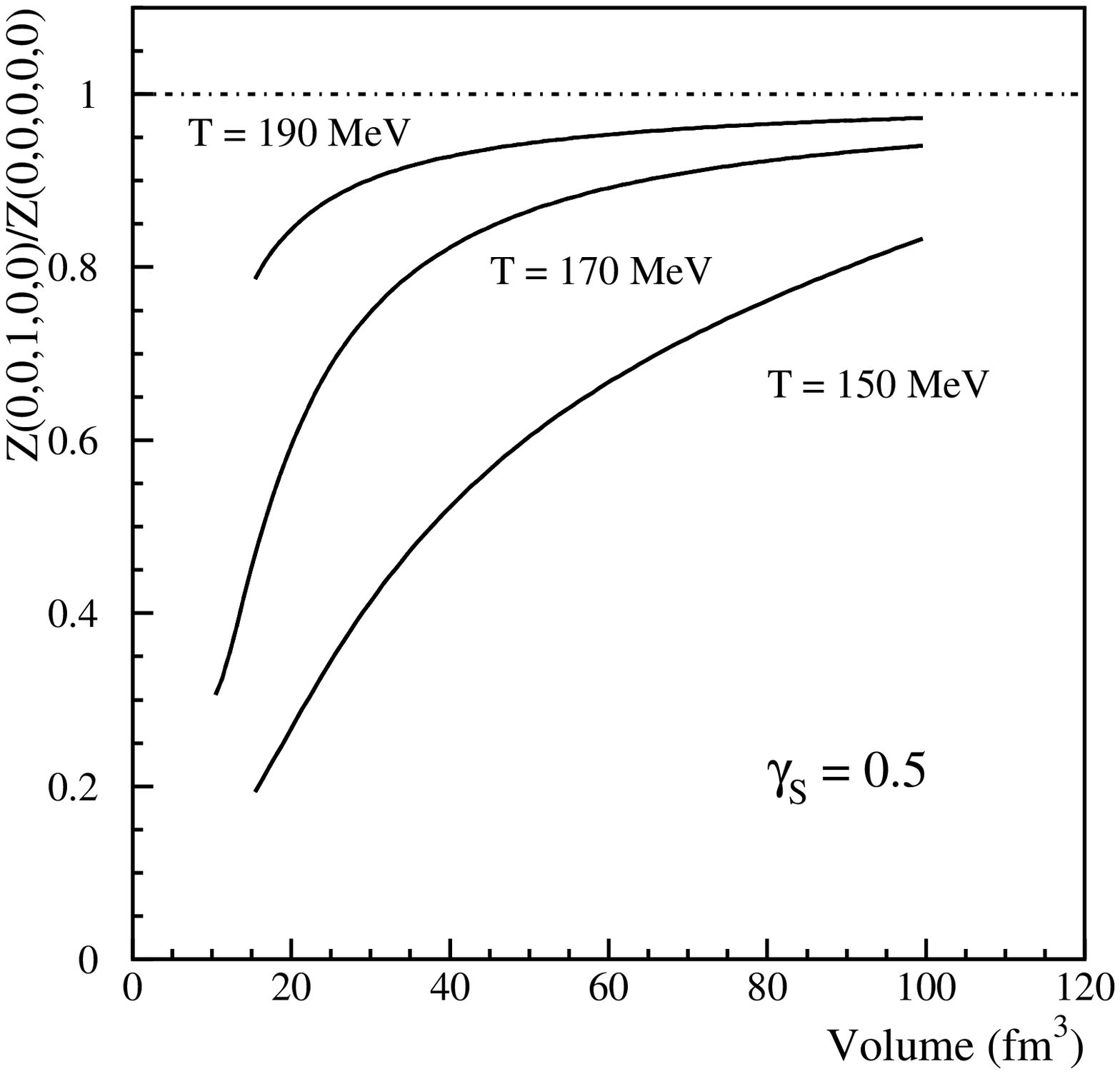,width=17cm}} 
\caption{}
\end{figure}

\newpage  
              
\begin{figure}[htbp]
\mbox{\epsfig{file=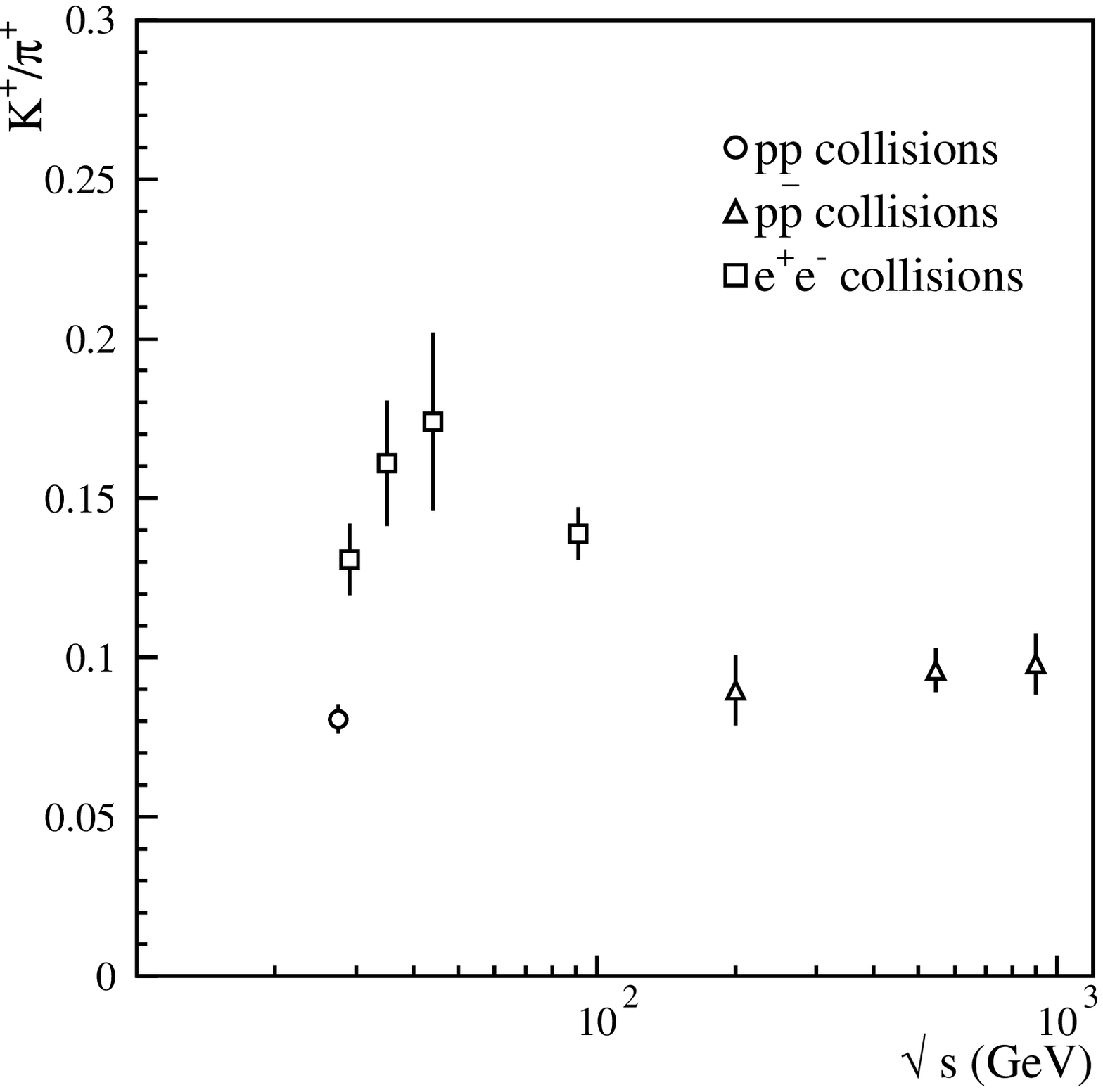,width=17cm}} 
\caption{}
\end{figure}

\end{document}